\documentclass[a4paper,twoside]{article}
\usepackage{geometry}
\usepackage{booktabs}
\usepackage{caption}	 

\usepackage{hyperref}
\usepackage{natbib}
\usepackage{graphicx}
\graphicspath{ {./} }

\usepackage{xcolor}
\usepackage{booktabs}

\usepackage{fancyhdr}
 
\makeatletter
\def\maxwidth{\ifdim\Gin@nat@width>\linewidth\linewidth
\else\Gin@nat@width\fi}
\makeatother
\let\Oldincludegraphics\includegraphics
\renewcommand{\includegraphics}[1]{\Oldincludegraphics[width=\maxwidth]{#1}}

\author{
			Wouter Groeneveld \\
		KU Leuven \\
		Leuven, Belgium\\
		\texttt{wouter.groeneveld@kuleuven.be}
		\and
			Hans Jacobs \\
		Prato \\
		Hasselt, Belgium\\
		\texttt{hans.jacobs@prato.be}
		\and
			Joost Vennekens \\
		KU Leuven \\
		Leuven, Belgium\\
		\texttt{joost.vennekens@kuleuven.be}
		\and
			Kris Aerts \\
		KU Leuven \\
		Leuven, Belgium\\
		\texttt{kris.aerts@kuleuven.be}
		\and
	}
\date{}

\begin{document}
\newcommand*\rot{\rotatebox{90}}

	\title{Non-cognitive abilities of exceptional software engineers: a Delphi
study}

\maketitle

\begin{abstract}
Important building blocks of software engineering concepts are without a
doubt technical. During the last decade, research and practical interest
for non-technicalities has grown, revealing the building blocks to be
various skills and abilities beside pure technical knowledge. Multiple
attempts to categorise these blocks have been made, but so far little
international studies have been performed that identify skills by asking
experts from both the industrial and academic world: \emph{which
abilities are needed for a developer to excel in the software
engineering industry}? To answer this question, we performed a Delphi
study, inviting 36 experts from 11 different countries world-wide,
affiliated with 21 internationally renowned institutions. This study
presents the 55 identified and ranked skills as classified in four major
areas: communicative skills (empathy, actively listening, etc.),
collaborative skills (sharing responsibility, learning from each other,
etc.), problem solving skills (verifying assumptions, solution-oriented
thinking, etc.), and personal skills (curiosity, being open to ideas,
etc.), of which a comparison has been made between opinions of technical
experts, business experts, and academics. We hope this work inspires
educators and practitioners to adjust their training programs,
mitigating the gap between the industry and the academic world.
\end{abstract}

\hypertarget{introduction}{%
\section{INTRODUCTION}\label{introduction}}

Knowledge of software development is becoming more and more important,
as shortcomings in the software engineering workforce require companies
to come up with creative solutions such as coding boot-camps, to
initiate candidates into the wonderful world of programming. However, to
be successful as a developer, it no longer suffices to be technically
proficient \citep{garousi2019closing}. There is still no agreement on
what separates a great developer from a good one, even if both the
academic and industrial world are starting to acknowledge the need for
something more besides cognitive knowledge, however good this might be
\citep{capretz2017soft}. Software is first and foremost created by
people, for people, hinting on the need for so-called `\emph{soft
skills}', or, more generally, `\emph{non-cognitive abilities}', defined
as the subset of abilities not related to technical skills.

In a previous literature study, we identified which non-cognitive
abilities are perceived as important to educators, and how these are
currently being taught \citep{groeneveld2019lit}. To complement these
findings and to shed more light on possible important skills missed by
the review, a Delphi study was performed, gathering data from software
engineering experts within both the academic and industrial world. With
this study, we aimed to identify and rank abilities besides the obvious
cognitive knowledge, perceived as a requirement for people concerned
with the technical facets of the software development process. We asked
ourselves, and therefore the experts who are more suitable to answer
this, the following questions:

\begin{quote}
\texttt{Q1}: \emph{Which non-cognitive abilities are needed for a
\mbox{developer}, to excel in the software engineering industry?}

\texttt{Q2}: \emph{What is the relative importance of these abilities?}

\texttt{Q3}: \emph{Does the opinion of industry experts differ from
academics, and if so, in what way?}
\end{quote}

The remainder of this paper is divided into the following sections.
Section 2 describes related work, while section 3 clarifies the process
we have utilized. Next, in section 4 and 5, we present and discuss the
results, followed by possible limitations of the study in section 6. The
last section, part 7, concludes this work.

\hypertarget{related-work}{%
\section{RELATED WORK}\label{related-work}}

`\emph{What makes a great software engineer?}' is the central question
Li et al.~answer by interviewing engineers at Microsoft
\citep{li2015makes}. Li identified a set of 53 general attributes of
great engineers, categorized into `\emph{personal characteristics}'
(improving, passionate, open-minded, etc.), `\emph{decision making}'
(knowledge-oriented), `\emph{teammates}' (mentoring, asking for help,
honest, etc.), and `\emph{software product}' (elegant, creative,
evolving, etc.). Even if the interviewed engineers come from different
divisions, the identified attributes still originate from only one
company. While the diverse set of attributes certainly help to
understand what affects the success of software projects, the lack of
prioritization renders it difficult to translate this list into concrete
recommendations for education.

A lot of similar investigations exist in literature, but few seem to
primarily target non-cognitive abilities. Some researchers look for
industrial requirements by analyzing job ads, risking the inclusion of
buzz-words and absence of implicitly required skills
\citep{stevens2016industry}. Ad content analysis and surveys are popular
tools used to research trends and identify (IT) skills. However, Delphi
studies usually offer more depth in terms of identification of a single
concept, such as critical skills for managing IT projects
\citep{keil2013understanding}, or soft competencies in requirements
engineering \citep{holtkamp2015soft}. Our aim with this study is the
same, focusing instead on modern software development.

Another Delphi study was conducted by Wynekoop et al.~in 2000, to
identify the traits of top performing software developers. This study
revealed the most important high-level traits to be geared towards
cognitivism \citep{wynekoop2000investigating}. Of course, within a
period of 20 years, the software development world has evolved
considerably, as the results of this study indicate.

Groeneveld et al.~reviewed literature on teaching non-technical skills
in software engineering education \citep{groeneveld2019lit}. They
defined communication, teamwork/dynamics, self-reflection, and conflict
resolution as most popular skills to teach students, while creativity,
ethics, and empathy dangle at the bottom of the priority list.

So far, we have not found a study that includes both the opinion of
people from the industry and people from the academic world. It will be
interesting to see what both parties have to say. As is the case with
most related studies, high-level discovered traits such as `\emph{good
interpersonal skills}', `\emph{team oriented}', `\emph{leadership
skills}', and `\emph{technical proficiency}' lack depth, serving as a
whole category rather than an individual trait. To mitigate this
shortcoming, we made sure to break down general abilities into multiple
specific low-level ones, as revealed in section 3.

\hypertarget{methodology}{%
\section{METHODOLOGY}\label{methodology}}

To be able to answer our research questions without merely relying on
analysis of existing data, we needed the help from industry veterans.
Since simply identifying skills would not sufficiently answer our
questions, we opted for the Delphi approach, of which we have applied
the most conventional implementation, collecting opinions of different
groups called \emph{panels}. The Delphi technique is an effective tool
to reach a group consensus on a given topic, where the opinion of
experts is used for decision making \citep{okoli2004delphi}.

As stated by Okoli et al., expert selection is one of the most important
aspects in a Delphi study \citep{okoli2004delphi}. We adhered to their
proposed participant selection model by identifying relevant
stakeholders, carefully reviewing their expertise based on information
at hand, such as blogs, CVs, social media, and papers, categorizing the
identified experts into separate stakeholder panels, and inviting
experts until a pre-set limit of 10 to 18 members per panel is reached.

Participating in a Delphi study requires a hefty commitment, since there
are several rounds to go through, with the first open-ended round being
the most cognitive demanding. Because of that, the availability of
information regarding the process and a correct formulation of questions
were given special attention. Potential panel members were contacted and
asked to participate to reduce drop-out rates during the study.

We invited experts from three different fields, as visible in Table
\ref{invites}: (T) \emph{technical experts} (software developers,
software architects, technical coaches, etc), (B) \emph{business
experts} (development and HR managers, agile coaches, analysts, etc),
and (A) \emph{academics}. The first two groups consist of people in
daily close contact with software engineering, while the last group
contains researchers and/or educators with experience in the same field.

\begin{table}[h!]
\captionsetup{belowskip=-10pt}
\centering
\caption{Panel groups, response rates, and mean duration in minutes for each phase.\label{invites}\label{table:Panel_groups,_respon}}
\begin{tabular}{l r r r r}
\hline
Round & Dur. & Panel T & Panel B & Panel A\\ [0.5ex]
\hline
\hline
Invites &  & 35 & 31 & 80\tabularnewline
Members &  & 14 (40\%) & 12 (39\%) & 10 (12\%)\tabularnewline
\hline
\#1 & 35m & 11 (79\%) & 11 (92\%) & 10 (100\%)\tabularnewline
\#2 & 6m & 11 (79\%) & 8 (67\%) & 7 (70\%)\tabularnewline
\#3 & 6m & 9 (64\%) & 10 (83\%) & 9 (90\%)\tabularnewline
\#4.1 & 10m & 9 (64\%) & 9 (75\%) & 8 (80\%)\tabularnewline
\#4.2 & 6m & 9 (64\%) & 8 (67\%) & 9 (90\%)\tabularnewline
\hline\end{tabular}
\end{table}

Our aim at answering the research questions was to deliberately not
limit ourselves to collecting responses from domestic software
engineers, but to also include opinions of as many widely recognized
experts as possible from well-known international companies. Hence the
choice of distributing the survey through email. This choice comes with
the unfortunate downside of initial low response rates combined with a
possible drop-out between rounds.

\begin{figure}[h!]
  \captionsetup{belowskip=-5pt}
  \centering
  \includegraphics{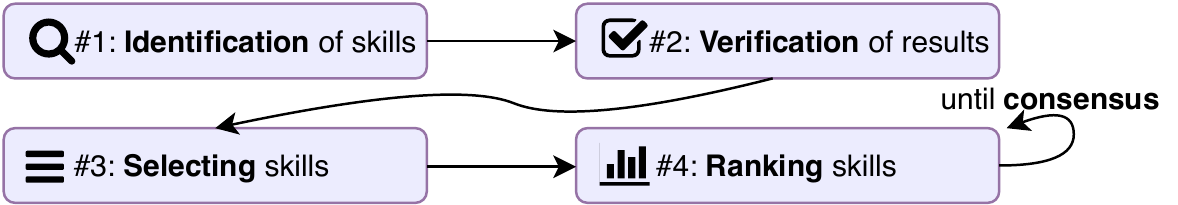}
  \caption{Delphi survey structure. \label{fig:delphi}}
\end{figure}

For this study, the Qualtrics web-based survey system was used.
Information was gathered as follows: starting with an open‐ended
questionnaire (1), after data processing and grouping, we ask
participants to verify and potentially revise their answers (2),
continuing with the selection of the top x most important skills (3), to
finish the survey with a series of ranking iterations (4), until a
consensus is reached or two iterations have passed. We will briefly
discuss each step below, as shown in Figure \ref{fig:delphi}.

\emph{1. Identification of skills}: The first questionnaire is the most
crucial one, designed as an open-ended digital brainstorming session.
Participants are asked to answer our first question by listing at least
\emph{five abilities} they think are important. Because we expected this
to be answered with typical high-level keywords such as `communication',
`teamwork', and `collaboration', we asked two more questions. First,
respondents were required to include a brief explanation (2 to 3
sentences) of \emph{why} these are important to them. Next, they had to
imagine a developer (possibly themselves) working with a colleague on a
problem, specifying \emph{what behaviour} they observe that shows this
colleague has the qualities answered in the first question. This extra
data made it possible for us to move from a general keyword to a
specific ability. In order to avoid confusion regarding the terms used
in the questions, we provided a definition of `abilities',
`non-cognitive abilities', and `developers', as consultable at
\mbox{\url{https://people.cs.kuleuven.be/~wouter.groeneveld/delphi}}.

\emph{2. Verification of results}: To ensure the correctness of coding
the open responses from round 1, we recruited the help of a seasoned
software developer (second author), who was familiar with the problem at
hand, but did not participate in the survey. He analyzed two thirds of
the anonymized raw data, creating his own set of categories. After
comparing and correcting the results, a final run through all the
responses resulted in the finalized list of identified abilities. After
this coding step, the next questionnaire asks respondents to identify
their answers, and if needed, provide further suggestions. This step in
the Delphi process verifies the correctness of our coding system, and
gives panel members a chance to add to their previous answer. When
discussing the results between authors, duplicates were eliminated, so
that the original wording of the respondents was retained as much as
possible.

\emph{3. Selecting most important skills}: Now that the complete list of
abilities has been identified, we asked participants to select \emph{at
least 10 factors} they consider to be the most important skills for a
developer to excel in the software engineering industry. In this round,
categories have been removed, and the order of all choices has been
fully randomized. It is important to notice that from round three and
up, the list of abilities will be narrowed down and ranked from within
each panel, while the first two rounds were conducted jointly by all
panels.

\emph{4. Ranking skills}: To reduce the ordered skill set from round 3
into a manageable size for each panel, only selected skills by at least
one member of this panel will remain for respondents to rank by
importance, using a drag-and-drop interface. The last round has to be
repeated until a general consensus is reached, measured by calculating
\emph{Kendall's W} \textgreater{}= \texttt{0.7} (\texttt{1} = complete
agreement, \texttt{0} = no agreement) \citep{okoli2004delphi}. If
\emph{W} does not rise to the desired level within two iterations, the
process will stop anyway, as to not overburden panel members with
questionnaires. The end result of the Delphi study nets us three
separate lists of ranked, important non-cognitive attributes, identified
by each group of stakeholders.

The results in between rounds were not fed back to the participants,
except for the data needed to compose the next round. Everyone received
a summary of their own answer after each round. Panel members did not
know each other, but their identity was visible for the researchers.
Data from drop-outs was not excluded from subsequent rounds, and no
additional people were invited once the study started.

\hypertarget{results}{%
\section{RESULTS}\label{results}}

Panel members completed the survey from 11 different countries
world-wide, with 17 members from Belgium. Members came from 12 different
companies and 9 different learning institutions, including our own
university. 146 people have been invited to take part, with 36 positive
responses, leading to an invite response rate of 25\%, with 39\% from
the industry and merely 12\% from academia, of which 89\% participants
successfully completed the first round.

Several well-known authorities in the field of enterprise software
development took part, as well as (co-)authors of praised books. We made
sure not to invite more than four experts from the same company or
university, reducing the risk of creating homogeneous groups. Companies
involved differed in size, from small to internationally renowned. Thus,
we believe panel members to have accumulated sufficient experience to be
able to contribute to this research.

Table \ref{fulltable} contains the full list of identified abilities,
including a final ranking for each expert panel. Our aim was to code the
skills as detailed as possible (e.g.~\emph{adjust your communication to
the target audience}, \emph{visually communicate ideas}, \emph{speak
well in front of an audience}) instead of listing high-level skills
(e.g.~\emph{communication}, \emph{presentation skills},
\emph{leadership}). Skill categories closely related with cognition
(e.g.~\emph{intelligence}, \emph{critical} and \emph{analytical
thinking}) were also split up, while too technical answers were omitted
from subsequent rounds, as these are not relevant to answer the research
questions.

There was very little consensus within each panel during the first
ranking round: \emph{W} was calculated at \texttt{0.16}, \texttt{0.12},
and \texttt{0.16} for panel T, B, and A, respectively. After rearranging
skills according to the new mean ranks and providing a summarized
motivation as feedback, the second ranking round reached a much higher
agreement: \emph{W} reached \texttt{0.74}, \texttt{0.80}, and
\texttt{0.58}, proving to be sufficient to end the survey.

\hypertarget{discussion}{%
\section{DISCUSSION}\label{discussion}}

Some abilities are closely related (e.g.~\emph{sharing knowledge} and
\emph{coaching others}), but still sufficiently dissimilar to justify
separation. However, some participants showed difficulties understanding
these subtle contrasts while prioritizing and ranking them in round 3
and 4, contributing to a low consensus rate. We felt that digital
surveys did not offer enough room for discussion between panel members
to clear out any disagreements on precisely defining these terms,
despite our attempts to answer questions and disseminate information
between rounds.

In general, we noticed a clear pattern in the answers, that is reflected
in two groups of two categories: communicative skills (1) \&
collaborative skills (2), and problem solving skills (3) \& personal
skills (4). As one participant rightfully observes:

\small

\begin{quote}
`\emph{To me, software engineering is a balancing act between being able
to communicate with other people and solving problems.}'
\end{quote}

\normalsize

The first group focuses on getting along, while the last two categories
contain problem solving skills and engineering inventiveness. We will
briefly discuss each category, highlighting interesting findings,
continuing with a look at the degree of consensus for each expert panel.

\begin{table*}
\small
\begin{center}
\begin{tabular*}{\textwidth}{l @{\extracolsep{\fill}} lc|cc|cc|cc}
\\
\toprule
 & & \textbf{Occ.}
 & \multicolumn{6}{c}{\textbf{\color{lightgray}{Occ. round 3 -} \color{black}{Final rank}}}
\\
 & \textbf{The ability to...}
 & \textbf{\#}
 & \multicolumn{2}{c}{\textbf{T}}
 & \multicolumn{2}{c}{\textbf{B}}
 & \multicolumn{2}{c}{\textbf{A}}
\\
\midrule
  & adjust your communication to the target audience & 11
  & \color{lightgray}{4} & $16^{th}$
  & \multicolumn{2}{c|}{/}  
  & \color{lightgray}{5} & $12^{th}$
\\
  & actively listen & 9
  & \color{lightgray}{6} & $10^{th}$
  & \color{lightgray}{5} & $10^{th}$
  & \color{lightgray}{4} & $6^{th}$
\\
  & understand and engage with the people involved during development & 8
  & \multicolumn{2}{c|}{/}
  & \multicolumn{2}{c|}{/} 
  & \color{lightgray}{6} & $1^{st}$
\\
  & clearly express intentions to avoid misunderstandings (including in code) & 8
  & \multicolumn{2}{c|}{/} 
  & \multicolumn{2}{c|}{/}
  & \multicolumn{2}{c}{/}
\\
  & show respect by exercising patience & 6
  & \multicolumn{2}{c|}{/} 
  & \multicolumn{2}{c|}{/} 
  & \multicolumn{2}{c}{/}
\\
  & visually communicate ideas/problems (e.g. drawing on a whiteboard) & 4
  & \color{lightgray}{5} & $13^{th}$
  & \color{lightgray}{4} & $13^{th}$
  & \multicolumn{2}{c}{/}
\\
  & empathize with others by understanding others' viewpoints & 4
  & \color{lightgray}{4} & $7^{th}$
  & \multicolumn{2}{c|}{/} 
  & \color{lightgray}{4} & $4^{th}$
\\
  & be aware of, and to a certain extend control, own and others' emotions & 2
  & \multicolumn{2}{c|}{/} 
  & \multicolumn{2}{c|}{/}
  & \multicolumn{2}{c}{/}
\\
  & have the courage to raise concerns & 2
  & \color{lightgray}{4} & $15^{th}$
  & \multicolumn{2}{c|}{/}
  & \multicolumn{2}{c}{/}
\\
  & demonstrate maturity by admitting when you're wrong & 2
  & \multicolumn{2}{c|}{/} 
  & \multicolumn{2}{c|}{/} 
  & \multicolumn{2}{c}{/}
\\
  \rot{\rlap{\textbf{Communicative Skills (11)}}}
  & speak well in front of an audience & 1
  & \multicolumn{2}{c|}{/} 
  & \multicolumn{2}{c|}{/}
  & \multicolumn{2}{c}{/}
\\
\midrule
  & collaborate with others to achieve a shared goal & 10
  & \color{lightgray}{4} & $6^{th}$
  & \color{lightgray}{5} & $1^{st}$
  & \color{lightgray}{7} & $2^{nd}$
\\
  & learn from each other by sharing/evangelizing/teaching/... knowledge & 9
  & \multicolumn{2}{c|}{/}
  & \color{lightgray}{6} & $6^{th}$
  & \multicolumn{2}{c}{/}
\\
  & be accountable by showing, taking, and sharing responsibility & 8
  & \color{lightgray}{4} & $12^{th}$
  & \color{lightgray}{5} & $4^{th}$
  & \multicolumn{2}{c}{/}
\\
  & give and receive feedback & 8
  & \color{lightgray}{4} & $17^{th}$
  & \color{lightgray}{6} & $9^{th}$
  & \color{lightgray}{4} & $8^{th}$
\\
  & (pro-actively) ask for help when needed & 6
  & \multicolumn{2}{c|}{/} 
  & \color{lightgray}{5} & $12^{th}$
  & \color{lightgray}{4} & $13^{th}$
\\
  & (pro-actively) help others when they encounter a problem & 5 
  & \multicolumn{2}{c|}{/}
  & \multicolumn{2}{c|}{/}
  & \multicolumn{2}{c}{/}
\\
  & negotiate a solution that resolves tensions between stakeholders & 4
  & \multicolumn{2}{c|}{/} 
  & \color{lightgray}{5} & $14^{th}$
  & \multicolumn{2}{c}{/}
\\
  & show and build a sense of belonging to team- and companywide values & 3
  & \multicolumn{2}{c|}{/}
  & \multicolumn{2}{c|}{/}
  & \multicolumn{2}{c}{/}
\\
  & mentor/coach/train others by providing guidance & 3
  & \color{lightgray}{4} & $18^{th}$
  & \multicolumn{2}{c|}{/} 
  & \color{lightgray}{4} & $15^{th}$
\\
  & promote well-being by contributing to team spirit & 2
  & \multicolumn{2}{c|}{/} 
  & \multicolumn{2}{c|}{/}
  & \multicolumn{2}{c}{/}
\\
  & maintain a positive attitude that inspires others & 2
  & \multicolumn{2}{c|}{/} 
  & \color{lightgray}{4} & $7^{th}$
  & \multicolumn{2}{c}{/}
\\
  & practice discipline by acting according to established team rules & 2
  & \multicolumn{2}{c|}{/} 
  & \multicolumn{2}{c|}{/}
  & \multicolumn{2}{c}{/}
\\
  & refrain from frequently interrupting others & 2
  & \multicolumn{2}{c|}{/} 
  & \multicolumn{2}{c|}{/} 
  & \multicolumn{2}{c}{/}
\\
  & work autonomously within a group & 2
  & \multicolumn{2}{c|}{/} 
  & \multicolumn{2}{c|}{/}
  & \multicolumn{2}{c}{/}
\\
  & resolve conflicts by taking advantage of new perspectives & 2
  & \multicolumn{2}{c|}{/} 
  & \multicolumn{2}{c|}{/} 
  & \multicolumn{2}{c}{/}
\\
  \rot{\rlap{\textbf{Collaborative Skills (16)}}}
  & accept diversity in the workplace & 1
  & \multicolumn{2}{c|}{/} 
  & \multicolumn{2}{c|}{/} 
  & \multicolumn{2}{c}{/}
\\
\midrule
  & show perseverance by not giving up easily and keep on trying again & 11
  & \color{lightgray}{4} & $11^{th}$
  & \multicolumn{2}{c|}{/} 
  & \multicolumn{2}{c}{/}
\\
  & be creative by approaching a problem from different angles & 10 
  & \color{lightgray}{7} & $3^{rd}$
  & \color{lightgray}{4} & $8^{th}$
  & \color{lightgray}{4} & $9^{th}$
\\
  & show attention to detail & 4
  & \color{lightgray}{5} & $9^{th}$
  & \multicolumn{2}{c|}{/} 
  & \multicolumn{2}{c}{/}
\\
  & focus on complex problems for longer, uninterrupted periods & 3 
  & \multicolumn{2}{c|}{/} 
  & \multicolumn{2}{c|}{/} 
  & \multicolumn{2}{c}{/}
\\
  & systematically verify assumptions and validate results & 3
  & \color{lightgray}{5} & $1^{st}$
  & \multicolumn{2}{c|}{/} 
  & \color{lightgray}{5} & $5^{th}$
\\
  & reframe problems by observing the bigger picture & 3 
  & \color{lightgray}{7} & $8^{th}$
  & \multicolumn{2}{c|}{/} 
  & \multicolumn{2}{c}{/}
\\
  & draw on multiple sources and domains for ideas & 2 
  & \multicolumn{2}{c|}{/} 
  & \multicolumn{2}{c|}{/} 
  & \multicolumn{2}{c}{/}
\\
  & create a clean structure leading to predictability and reliability & 2 
  & \multicolumn{2}{c|}{/} 
  & \multicolumn{2}{c|}{/} 
  & \multicolumn{2}{c}{/}
\\
  & like the challenge of a difficult problem & 1 
  & \multicolumn{2}{c|}{/} 
  & \multicolumn{2}{c|}{/} 
  & \multicolumn{2}{c}{/}
\\
  & think in a solution-oriented way by favoring outcome over ego & 1
  & \color{lightgray}{4} & $14^{th}$
  & \color{lightgray}{7} & $2^{nd}$
  & \multicolumn{2}{c}{/}
\\
  & focus on efficiency and re-use instead of reinventing the wheel & 1 
  & \multicolumn{2}{c|}{/} 
  & \multicolumn{2}{c|}{/} 
  & \multicolumn{2}{c}{/}
\\
  & measure (own/others/project) progress and improvement & 1
  & \multicolumn{2}{c|}{/} 
  & \multicolumn{2}{c|}{/} 
  & \multicolumn{2}{c}{/}
\\
  \rot{\rlap{\textbf{Problem Solving Skills (13)}}}
  & know which actions to take under which circumstances & 1 
  & \multicolumn{2}{c|}{/} 
  & \multicolumn{2}{c|}{/} 
  & \multicolumn{2}{c}{/}
\\
\midrule
  & be curious by wondering why and how something works & 10 
  & \color{lightgray}{7} & $2^{nd}$
  & \multicolumn{2}{c|}{/} 
  & \color{lightgray}{5} & $10^{th}$
\\
  & devote oneself to continuous learning & 8
  & \color{lightgray}{8} & $4^{th}$
  & \color{lightgray}{7} & $3^{rd}$
  & \color{lightgray}{5} & $14^{th}$
\\
  & be open to the ideas of others & 7 
  & \color{lightgray}{7} & $5^{th}$
  & \color{lightgray}{4} & $5^{th}$
  & \color{lightgray}{6} & $3^{rd}$
\\
  & be a fast learner to keep up with the pace of new technologies & 7
  & \multicolumn{2}{c|}{/}
  & \multicolumn{2}{c|}{/} 
  & \color{lightgray}{4} & $16^{th}$
\\
  & be passionate in what you do & 5
  & \multicolumn{2}{c|}{/} 
  & \multicolumn{2}{c|}{/} 
  & \multicolumn{2}{c}{/}
\\
  & criticize oneself and others objectively and fairly & 4
  & \multicolumn{2}{c|}{/} 
  & \multicolumn{2}{c|}{/} 
  & \color{lightgray}{4} & $11^{th}$
\\
  & be flexible by quickly adapting to changes during development & 3
  & \multicolumn{2}{c|}{/} 
  & \multicolumn{2}{c|}{/} 
  & \color{lightgray}{5} & $7^{th}$
\\
  & be consistent by doing as you say and saying as you do & 2
  & \multicolumn{2}{c|}{/} 
  & \multicolumn{2}{c|}{/} 
  & \multicolumn{2}{c}{/}
\\
  & set goals and stick to them & 2 
  & \multicolumn{2}{c|}{/} 
  & \multicolumn{2}{c|}{/} 
  & \multicolumn{2}{c}{/}
\\
  & be adventurous by willing to take risks & 1
  & \multicolumn{2}{c|}{/} 
  & \multicolumn{2}{c|}{/} 
  & \multicolumn{2}{c}{/}
\\
  & carefully choose which new trends to follow and which to ignore & 1
  & \multicolumn{2}{c|}{/} 
  & \multicolumn{2}{c|}{/} 
  & \multicolumn{2}{c}{/}
\\
  & be stress-resilient & 1 
  & \multicolumn{2}{c|}{/} 
  & \multicolumn{2}{c|}{/} 
  & \multicolumn{2}{c}{/}
\\
  & to cope with and learn from failure & 1 
  & \multicolumn{2}{c|}{/} 
  & \color{lightgray}{4} & $11^{th}$
  & \multicolumn{2}{c}{/}
\\
  & develop a sense of pride in your work & 1 
  & \multicolumn{2}{c|}{/} 
  & \multicolumn{2}{c|}{/}
  & \multicolumn{2}{c}{/}
\\
  \rot{\rlap{\textbf{Personal Skills (15)}}}
  & leverage self-knowledge to continuously improve oneself & 1 
  & \multicolumn{2}{c|}{/}
  & \multicolumn{2}{c|}{/}
  & \multicolumn{2}{c}{/}
\\
\bottomrule
\end{tabular*}
\caption{\small The full list of 55 identified skills, according to technical experts (T), business experts (B), and academics (A), sorted by occurence in round 1. Occurences per panel from round 3 are indicated in grey. More information, including a comparative visualization, is available at \mbox{\url{https://people.cs.kuleuven.be/~wouter.groeneveld/delphi}}. \label{fulltable}}
\end{center}
\end{table*}

\hypertarget{communicative-skills}{%
\subsection{Communicative Skills}\label{communicative-skills}}

As mentioned in \citep{groeneveld2019lit}, terms such as `communication'
and `teamwork' are most common when investigating skills to teach
students. It is therefore no surprise that abilities such as
\emph{understanding and engaging with people involved}, \emph{adjusting
communication to the target audience}, and \emph{actively listening} are
reflected in the results. It is interesting to note that technical
experts selected 5 out of 11 communicative skills to be of great
importance, while business experts only selected 2, and academics 4.
\emph{Having the courage to raise concerns} was only picked by panel T
and \emph{understanding and engaging with the people involved} only by
panel A. Communication is also seen as a way to structure and guide the
flow of information within a group:

\small

\begin{quote}
`\emph{Teams usually have to find a way to communicate, to try out
several ideas. Communication also requires structure: a common system to
make notes and have weekly/daily meetings to discuss these notes and
decide how to continue.}'
\end{quote}

\normalsize

Frequently mentioned skills during the first round such as \emph{clearly
expressing intentions to avoid misunderstandings} (8 occurrences) and
\emph{showing respect by being patient} (6) were not deemed important
enough to make it to the final rankings.

\hypertarget{collaborative-skills}{%
\subsection{Collaborative Skills}\label{collaborative-skills}}

Being a good communicator is not enough: enterprise software development
also requires working together, \emph{in close collaboration to achieve
a shared goal}, the most popular skill in this category. It is
self-evident that communicative skills also work in favor of
collaboration. Therefore, some items are interchangeable between this
category and the first.

Business experts especially seem to appreciate these skills, deeming 7
out of 16 of them as important in round 3, with \emph{learning from each
other by sharing/evangelizing/teaching knowledge}, \emph{negotiating},
and \emph{maintaining a positive attitude} being skills not picked by
the other two panels. However, \emph{mentoring/coaching/training others}
was indeed picked, and might also be interpreted as `sharing knowledge'.
Skills that promote \emph{belongingness} have been highlighted by
several respondents, reporting that software development is a team sport
nowadays:

\small

\begin{quote}
'\emph{The ability to function in a team, to help with the norming,
forming and performing of a team is crucial. In the end it's the team
who is committed to the success of the software, so we need developers
who are committed to the team.'}
\end{quote}

\normalsize

The same holds true for \emph{negotiating}:

\small

\begin{quote}
`\emph{Software is usually written for non-technical customers and
stakeholders, for commercial reasons. There is always a tension between
technical constraints, commercial needs, the needs and desires of
different stakeholders etc. Successful software development involves
negotiating a solution that best resolves those tensions.}'
\end{quote}

\normalsize

However, compared to other items in the result set, these barely made it
to the final ranking. \emph{Asking for help when needed} is seen as more
important than \emph{refraining from interrupting others}, and
\emph{working autonomously} again was given less attention than heavy
team-oriented skills.

\hypertarget{problem-solving-skills}{%
\subsection{Problem Solving Skills}\label{problem-solving-skills}}

Besides getting along, a second central theme to the skill set of a good
software developer is the ability to identify and effectively solve
problems. The Software Engineering Body of Skills (SWEBOS) list by
Sedelmaier et al.~splits this category into \emph{consciousness of
problems} and \emph{competence to solve problems}
\citep{sedelmaier2014software}. Response data also shows that being able
to identify a problem is something different than being able to actually
solve it. It comes to no surprise that technical experts marked 6 out of
13 skills from this category as important. What did surprise us was the
low amount of these skills picked up by other panels.

\emph{Showing perseverance}, the most popular skill (11 occurrences),
was only chosen by panel T, as did \emph{attention to detail} (4), and
\emph{reframing problems by observing the bigger picture} (3). We feel
that the ability to \emph{focus on complex problems for longer,
uninterrupted hours}, although mentioned (3), is underrated, as this
kind of `deep work' or `flow' has been proven to aid in problem solving
\citep{newport2016deep}. When it comes to software development,
`\emph{the devil is in the details}', as one respondent said:

\small

\begin{quote}
`\emph{When telling a computer what to do, the devil is in the details.
I have never known a good programmer who doesn't pay careful attention
to the details of the problem, the code, and the process.}'
\end{quote}

\normalsize

Many skills, including \emph{creativity}, can be amplified in the
presence of other abilities. They are connected in a systemic way that
is difficult to identify in a Delphi study, apart from analyzing open
ended question responses, such as this one:

\small

\begin{quote}
`\emph{The ability to think of new and creative ways to do things is
essential for any kind of problem solving. but it is particularly
important in software where complexity requires good design. Creativity
includes curiosity, questioning, exploring multiple paths, and
converging on a design.}'
\end{quote}

\normalsize

\hypertarget{personal-skills}{%
\subsection{Personal Skills}\label{personal-skills}}

The last category represents characteristic attributes of individuals,
as a combination between willingness and the ability to use a skill is
required to succeed. These dispositions include \emph{curiosity},
\emph{being open to ideas of others}, \emph{self-criticism}, and
intrinsic motivation such as \emph{being passionate}, which are all
consistently reported in critical thinking research
\citep{facione2000disposition}. Some items from the previous category
can also be placed within the context of personal attributes.

For academics, this category is the most important one, with 6 out of 16
skills that made it to the final ranking, while other panels only
include 3. While \emph{being a fast learner to keep up with the pace of
new technologies} (7 occurrences), \emph{criticize oneself and others}
(4), and \emph{being flexible by quickly adapting to changes} (3) are
popular choices, they failed to stay on target except in panel A.
\emph{Coping with and learning from failure} is an ability picked only
by business experts. Developers do acknowledge the need for resilience,
as one panel member stated:

\small

\begin{quote}
`\emph{Things will go wrong, you will make mistakes, you will
misunderstand requirements or technology. You need to be able to not
just cope with, but learn from that failure.}'
\end{quote}

\normalsize

This is also called `\emph{creating a safe haven}' by
\citep{li2015makes}, where engineers can learn and improve from mistakes
without negative consequences.

\hypertarget{expert-panel-opinions}{%
\subsection{Expert panel opinions}\label{expert-panel-opinions}}

\begin{figure}[h!]
\captionsetup{belowskip=-5pt}
\centering
\includegraphics{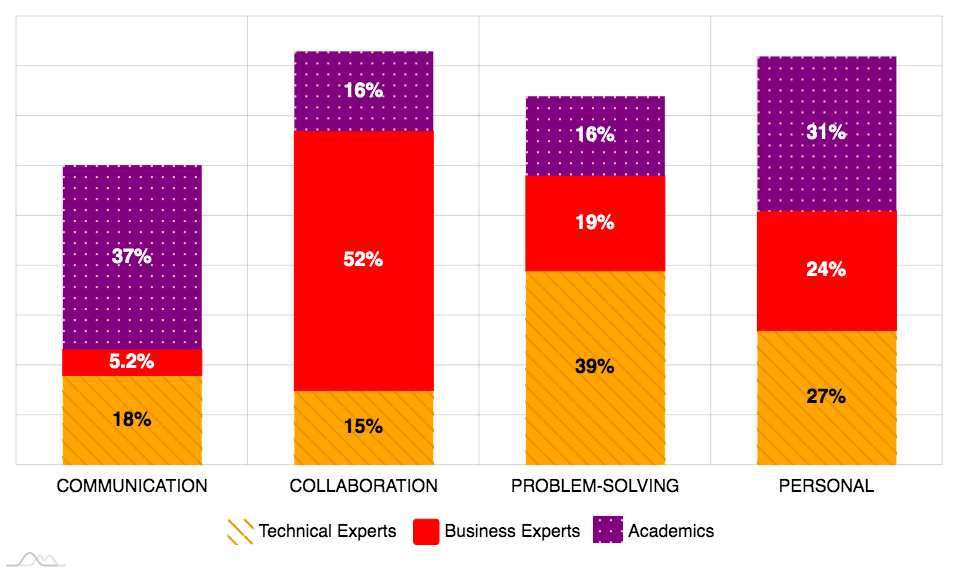}
\caption{A visual comparison between expert panel opinions, divided in
the four skill categories. \label{fig:comparison}}
\end{figure}

Perhaps unsurprisingly, industry and academic experts reached a
significant difference in opinion, as also reported by Eskandari et al.,
in a Delphi study to suggest engineering curriculum enhancements
\citep{eskandari2007enhancing}. Figure \ref{fig:comparison} visualises
these differences by comparing the normalized sum of ranked skills as
weights for each category.

\emph{Technical experts} (panel T) claim \emph{systematically verifying
assumptions} (\#1) is one of the most important abilities to be
successful, together with \emph{curiousness} (\#2) and \emph{creativity}
(\#3). Without \emph{continuously improving} (\#4) and learning, any
necessary non-cognitive skill becomes unsustainable. Problem solving
skills and personal skills matter most.

\emph{Business experts} (panel B) hold \emph{collaborating to achieve a
shared goal} (\#1) in high regard, next to \emph{thinking in a
solution-oriented way} (\#2). After all, the most important deliverable
for a software developer is working software, amplified by shared
success that may involve some personal compromises \citep{li2015makes}.
\emph{Continuously improving} (\#3) comes in third. There are some
similarities between business and technical experts. A good balance
between problem solving, personal skills, and collaborative skills is
required, with a heavy emphasis on collaboration, as visible in Figure
\ref{fig:comparison}.

\emph{Academics} (panel A) consider \emph{understanding and engaging
with people involved} (\#1) to be an important aspect, together with
\emph{collaborating to achieve a shared goal} (\#2) and \emph{being open
to the ideas of others} (\#3). They conclude that personal skills and
communication are of the utmost importance. The difference in opinion
about communicative skills between academics (37\%) and business experts
(5\%) is striking.

A possible explanation for the heavy emphasis on problem solving for
technical experts could be the tangible and immediate usability, while
more abstract personal characteristics that are preferred by academics
help with the norming and forming of those practical skills. However,
the industry seems to expect ready-to-use practical problem solving
skills \citep{bailey2014non}. This may depend on your professional
experience, as one respondent points out:

\small

\begin{quote}
`\emph{I found it very difficult to rank the skills by importance.
Relative importance can depend on where you are on your career.}'
\end{quote}

\normalsize

This is another reason why reaching a certain degree of consensus within
each group was not easily achievable.

\hypertarget{limitations}{%
\section{LIMITATIONS}\label{limitations}}

By carefully following the steps explained in section 3, and by
comparing our data with existing research, we are convinced that our
results can answer \texttt{Q1} and \texttt{Q3} safely. However, since a
Delphi study never reaches any statistical relevance, the answer to
\texttt{Q2} regarding relative importance of each skill might vary,
depending on the formation of the panel. Also, since the results are
highly contextual, ranking them might not always be very relevant.

As mentioned before, depending on where you are in your career, certain
skills might be more important to you than others. Also, depending on
the interpretation of the questions or the abilities themselves, answers
might vary. It is very difficult to facilitate a heated debate using
only a digital survey.

This could be mitigated by performing additional focus groups using the
results from this study as input, or by repeating round 3/4 in Figure
\ref{fig:delphi} with a new set of experts, until the outcome remains
steady. Regional and cultural differences might further complicate
things. Generalization was, however, never the goal for this research.
We believe that our systematic procedure yielded interesting and
relevant findings.

So far, we have been using the term `non-cognitive skills' to identify
the group of abilities. We asked participants to come up with a more
fitting alternative, given their selection of skills. The answers vary
and are listed at our website. While no single answer seems fitting
enough to harbor each category and result from Table \ref{fulltable}, it
is clear to us that `non-cognitive' is the wrong word, as one
participant elaborates:

\small

\begin{quote}
`\emph{Non-cognitive seems not only incorrect but derogatory. All
non-technical skills are cognitive, require practice, consideration and
deliberate exercise to improve.}'
\end{quote}

\normalsize

\hypertarget{conclusion}{%
\section{CONCLUSION}\label{conclusion}}

The 36 panel members of the Delphi study presented in this paper agreed
upon important abilities needed for a developer to excel in the software
engineering industry. By visualizing and comparing results, we have
uncovered interesting differences in opinion between panels. These
findings will hopefully help to close the gap between industry and
higher education. Future work might dig deeper into understanding these
dissimilarities.

The legitimate remark regarding your personal career stance raises the
question of which skills are absolutely essential for a starting
developer, and which skills will be developed as your experience grows.
The skill categories and voting trends visible in Table \ref{fulltable}
are more interesting for further investigation than the level of
agreement within the ranking rounds. It is important to ask ourselves
which skills from the list should be taught to students, and which
skills they will develop themselves as practitioners.

To further reconcile industry and academia, our future work will involve
zooming in on underexposed abilities that were considered important by
either party, and to investigate how to teach and evaluate these
abilities in higher education.

\hypertarget{acknowledgements}{%
\section{ACKNOWLEDGEMENTS}\label{acknowledgements}}

We wish to thank all survey respondents for their valuable contributions
to this research, who despite the annoying reminder emails and long
questionnaires, still managed to find time and effort to participate.
Thank you all very much!

\bibliographystyle{ACM-Reference-Format}
\bibliography{delphi.bib}

\end{document}